\documentclass[final]{aipproc}

\layoutstyle{8x11double}
\newcommand{\beq}{\begin{equation}}
\newcommand{\eq}{\end{equation}}
\newcommand{\bed}{\begin{displaymath}}
\newcommand{\ed}{\end{displaymath}}
\newcommand{\reff}[1]{\mbox{Fig.\hspace{2pt}\ref{#1}}}

\newcommand{\sgra}{\mbox{Sgr~$\mathrm{A}^{*}$}\ }

\newcommand{\rg}{r_{\mathrm{g}}}

\newcommand{\mbh}{M_{\mathrm{bh}}}

\newcommand{\parc}[2]{\frac{\partial #1}{\partial #2}}

\usepackage{amsmath}
\usepackage{amssymb}
\usepackage[mathcal]{eucal}

\begin{document}
\title{Tidal effects in the vicinity of a black hole}
\classification{ 98.35.Jk, 98.35.Mp, 98.54.-h, 98.54.Cm, 04.70.-s }
\keywords{Galaxy: nucleus - galaxies: active - black hole physics}
\author{Andrej \v Cade\v z}{address={Faculty of Mathematics and Physics, University of Ljubljana, Jadranska 19, 1000 Ljubljana, Slovenia} }
\author{Uro\v s Kosti\' c}{address={Faculty of Mathematics and Physics, University of Ljubljana, Jadranska 19, 1000 Ljubljana, Slovenia} }
\author{Massimo Calvani}{address={INAF - Astronomical Observatory of Padova, Vicolo Osservatorio 5, 35122 Padova, Italy}}
\begin{abstract}
The discovery that the Galactic centre emits flares at various wavelengths represents a puzzle concerning their origin, but at the same time it is a relevant opportunity to investigate the environment of the nearest super-massive black hole. In this paper we shall review some of our recent results concerning the tidal evolution of the orbits of low mass satellites around black holes, and the tidal effect during their in-fall.  We show that tidal interaction can offer an explanation for transient phenomena like near infra-red and X-ray flares from Sgr A*.
\end{abstract}
\maketitle
%
%
\section{Introduction}
It is nowadays widely accepted that most galaxies, if not all, contain a super-massive black hole at their centre. How these black holes are formed and how they are related to galaxy formation is still an open issue, see e.g. \cite{Rees:2006}. The nearest super-massive black hole (SMBH) is located at the centre of our Galaxy, Sgr A*.  Stellar orbits determinations \cite{Ghez1:2005} show that a central dark mass of $(3.7\pm 0.2)\times 10^6 [R_0/(8\, {\rm kpc})]^3 M_\odot$ (with an updated value of $(3.61\pm 0.32)
\times 10^6 M_\odot$ from observations with SINFONI \cite{Eisenhauer:2005}) is confined within a radius of 45 AU. The proximity of the Galactic centre ($\approx 8$ kpc) provides us a unique opportunity to study in detail the environment of massive black holes (see e.g.\ \cite{Alexander:2005} that contains an exhaustive description of the Galactic centre), and possibly of their formation. Due to its proximity, Sgr A* is the only galactic nucleus that can be investigated using VLBI on sub-AU linear scales, and observations in 2007 have confirmed structure in Sgr A* on scales of just a few Schwarzschild radii \cite{Doeleman:2008}.

The recent detection of rapid flare activity coming from Sgr A* represents additional strong evidence for the presence of a massive black hole. Their time scales strongly suggest motion only a few Schwarzschild radii away from the central black hole. If such flares are uncorrelated single events of accretion, we believe that the most plausible candidate for their energy source is the tidal part of the gravitational potential energy. We suggest that Galactic flares are produced by the final accretion of a single dense object, like an asteroid or a comet, with a mass of $\approx 10^{20}$g. We explore this scenario, and we show that the light curve of a flare can be deduced from dynamical properties of geodesic orbits around black holes and that it only weakly depends on the physical properties of the source.
\section{Flares from the Galactic centre}
Flares from Sgr A* have been detected in X-rays (Chandra and XMM-Newton satellites) and infra-red (VLT adaptive optics imager NACO, SINFONI infrared adaptive optics integral spectrometer on the ESO VLT). The description of this flaring activity can be found in several papers, e.g. \cite{Baganoff:2001, Genzel:2003, Ghez:2004, Belanger:2006, EckartB:2006, MeyerA:2006}. These flares are modulated on a short time-scale with average periods of $\approx 20$ minutes, and multi-wavelength campaigns found that the time lag between X-ray and NIR flare emission is very small, strongly suggesting a common physical origin, see e.g. \cite{Eckart:2004, EckartA:2006, Yusef-Zadeh:2006, Trippe:2007}.

The main features of the observed flares are: a) duration: $\approx 1/4$h -- 2h; b) frequency: few events per day; c) energy release: of the order of $10^{35}$ erg/sec; d) linear polarization (NIR, radio): of the order of few to few ten percent, with stable polarization angle; e) strong quasi-periodic modulation on time scales of minutes.

Several different models were proposed to explain these phenomena: disk instabilities (e.g. \cite{Tagger:2006, Yuan:2006, Falanga:2008, Eckart:2008}), star-disk interaction (e.g. \cite{Nayakshin:2004}), expanding hot blobs (e.g. \cite{Yusef-Zadeh:2006, Marrone:2008}), hot spot/ring models (e.g. \cite{Trippe:2007, Eckart:2004, EckartA:2006, Belanger:2006, MeyerA:2006, MeyerB:2006, Broderick:2006, Zamaninasab:2008}). Here we do not want to compare and contrast the above proposed model. Our aim is to propose a scenario in which  Galactic flares are produced by the final accretion of a single dense object, like an asteroid or a comet. The main motivations of this approach are: a) the
time-scales of flares strongly suggest motion only a few Schwarzschild radii away from the black hole; b) the energy release during these flares corresponds to the source mass of the order of $10^{20}$g, see e.g.\ \cite{Genzel:2003}; c) it seems reasonable to assume that stars at the Galactic centre are surrounded by planets and by other small orbiting bodies, like asteroids and comets (hereafter: LMS, low-mass satellites); d) tidal evolution of the orbits of LMS will lead to capturing orbits; e) tidal effects in the vicinity of a black hole can release a significant amount of energy.

Asteroid-like objects fit both observational facts (time-scales and energetics): they are of the right mass and they are tidally disrupted closer to the black hole than gaseous blobs. In active galaxies, such events would be outshined, but in the inactive centre of our Galaxy, it is possible to observe them.

In the following Sections we shall expand on the above motivations. Most of this work is based on our recent papers, e.g. \cite{Gomboc:2005, Cadez:2008, Kostic:2008, Kostic:2009}, where a thorough discussion and more details can be found.
\section{LMS at the Galactic centre}
According to \cite{Cochran:1995}, the Edgeworth-Kuiper belt of our Solar System may still (4.5 billion years after its creation) contain as many as  $2\times 10^8$ objects of radii $\lesssim 10$ km. It appears reasonable to assume that stars at the Galactic centre are surrounded by planets and by other small orbiting bodies (LMS), like asteroids and comets,  that may be stripped off their parent stars by tidal interaction, while approaching the black hole. Therefore, we expect that there must be a considerable number of solid objects that cluster the Galactic centre.

Little is known about dynamics that determines the fate of such low mass satellites. However, some hints come from the investigation of core dynamics at the Galactic centre \cite{Berukoff:2006}; from the investigation of mechanisms that may cause a star to move inwards into a massive black hole, in connection with possible detection of gravitational waves with LISA, see \cite{Hopman:2006} and references therein; from detailed dynamical studies of the evolution of the Solar System, see e.g. \cite{Dermott:1988, Faber:2005, Mudryck:2006}.

One might expect to find stellar system satellites all the way down to the black hole at the Galactic centre, with a fair proportion of them on low-periastron, highly eccentric orbits. A significant work will be done by tidal forces near the low periastron. This work partially dissipates as heat, and as a result it lowers orbital energy and may start significant tidal evolution of the orbit, see \cite{Cadez:2008}. When periastron crossing frequency and fundamental quadrupole frequency of the object are the same, that is at resonance, tidal interaction is strongest. In this context, one must consider two classes of LMS: those that are gravity dominated and those that are solid state forces dominated.

For gravity dominated objects the fundamental frequency is $\nu_{\rm g} \approx 2 \sqrt{G \rho / 3\pi}$, while for solid state dominated objects it is $\nu_{\rm s}=\frac{1}{4}c_{\rm s}/R$ ($\rho$ is the density of the body, $R$ its radius and $c_{\rm s}$ the speed of sound). Taking as typical values $c_{\rm s}\approx 5\ \rm{km/s}$ and $\rho\approx 5\ \rm{g\ cm^{-3}}$, we find that the radius dividing the two classes is about $1000\ \mathrm{km}$, i.e. the radius of the asteroid Ceres. Accordingly, all gravity dominated satellites must have about the same fundamental quadrupole frequency, that corresponds to a period of about 1 hour and all smaller satellites have shorter fundamental periods. Thus, gravity dominated satellites start a rapid, resonant tidal evolution when their periastron reaches $r_{\rm p}\approx 10\ r_{\rm g}$ ($r_{\rm g} = GM_{\rm bh}/c^2$), while solid state dominated bodies may undergo significant tidal evolution even closer to the black hole.

\begin{figure}
\includegraphics[width=\columnwidth]{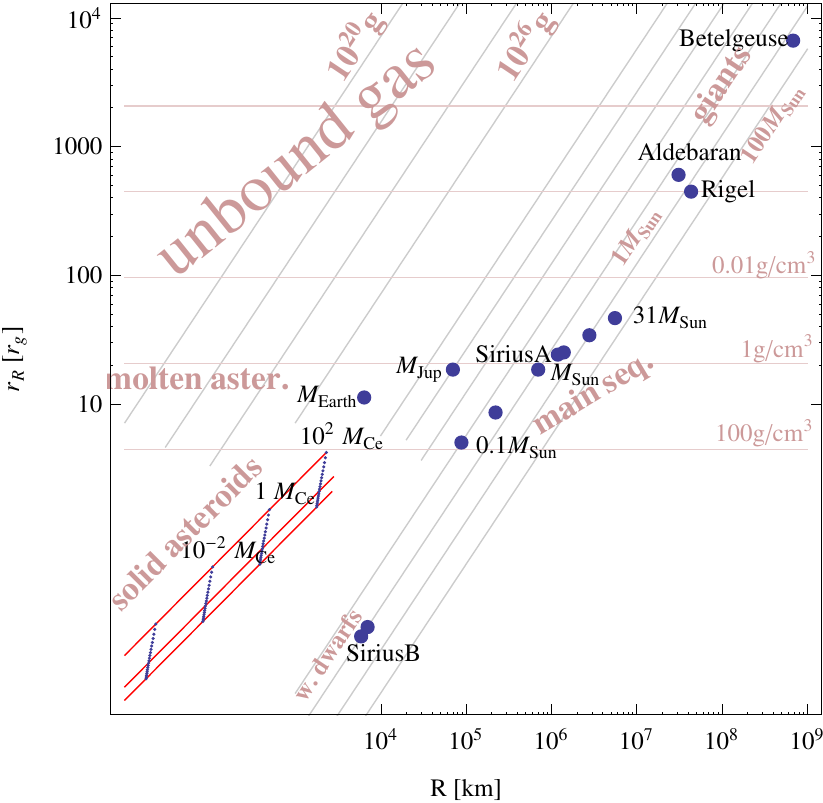}
  \caption{Astrophysical objects capable of releasing significant tidal energy before falling down the Galactic centre black hole. The positions of different celestial objects are also marked in this diagram. Note that the effective Roche radius for solid state dominated asteroids is below $2\ r_{\rm g}$, while the Roche radius for these objects is about $10\ r_{\rm g}$.}
\label{Fig:GCobjects}
\end{figure}

The difference between these two classes of objects is better understood introducing the Roche radius and the effective Roche radius. The effective Roche radius is the radius at which $\omega_{\rm q} T_{\rm f} = 1$, where $\omega_{\rm q}$ is the fundamental resonant frequency of quadrupole modes and $T_{\rm f}=(2 r_{\rm p}^3/ GM_{\rm bh})^{1/2}$ is the periastron crossing time \cite{Gomboc:2005}. In \reff{Fig:GCobjects} we show the Roche radius and the effective Roche radius as a function of object size, mass and density for different bodies expected to populate the Galactic centre. This graph accentuates the very particular position of asteroids. With the exception of compact stars such as white dwarfs, neutron stars or stellar mass black holes, they are the only objects that can survive the tidal field of the black hole when held together by solid state forces. All other objects such as planets, main sequence stars and of course giants find themselves beyond the Roche radius above ten or a few ten gravitational radii. Note, however, that molten asteroids become gravity dominated. If melting occurs suddenly below $\sim 10\ r_{\mathrm{g}}$, the phase change results in a dramatic rapidly developing tidal disruption. We intend to show that why and how this may happen.
\section{Tidal evolution of LMS orbits}
Tidal effects on a star encountering a SMBH have been investigated by several authors, see e.g. \cite{Gomboc:2005, Brassart:2008} and references therein. However, tidal forces will not affect only the shape and the structure of the in-falling body, but are responsible for the evolution of the orbits and the rotation periods of the involved objects.

The first stages of tidal evolution of LMS orbits can be investigated using Hut's formalism described in \cite{Hut:1980, Hut:1981, Hut:1982}. This formalism is suitable for describing slow tides, i.e. tides that change slowly with respect to the frequency of the fundamental quadrupole modes of the object that tides are acting upon. In this case the relevant forces that exchange angular momentum and energy can accurately be calculated by approximating the deformed object adding two additional bulge masses $\mu$ on the surface of the star. In this case tidal evolution of orbits depends only on the parameter $\alpha=l_0 /s_0$, which is the ratio of orbital angular momentum of the system and spin of the secondary at equilibrium. Therefore, completely different binary systems with the same value of $\alpha$ behave in the same way. See \cite{Kostic:2008} for details.

We find that the evolution of orbits strongly depends on the initial value of $\tilde {\rm r}_{p,0}$, which is the ratio between the periastron distance ${\rm r}_{p}$ and the radius of the equilibrium circular orbit ${\rm a}_0$. If an orbit starts with $\tilde {\rm r}_{p,0}$ below some critical value, it will always end up in in-spiralling, regardless of the value of $\alpha$. Orbits starting above this critical value of $\tilde {\rm r}_{p,0}$ will either circularize, or become unbound.

\begin{figure}
\includegraphics[width=\columnwidth]{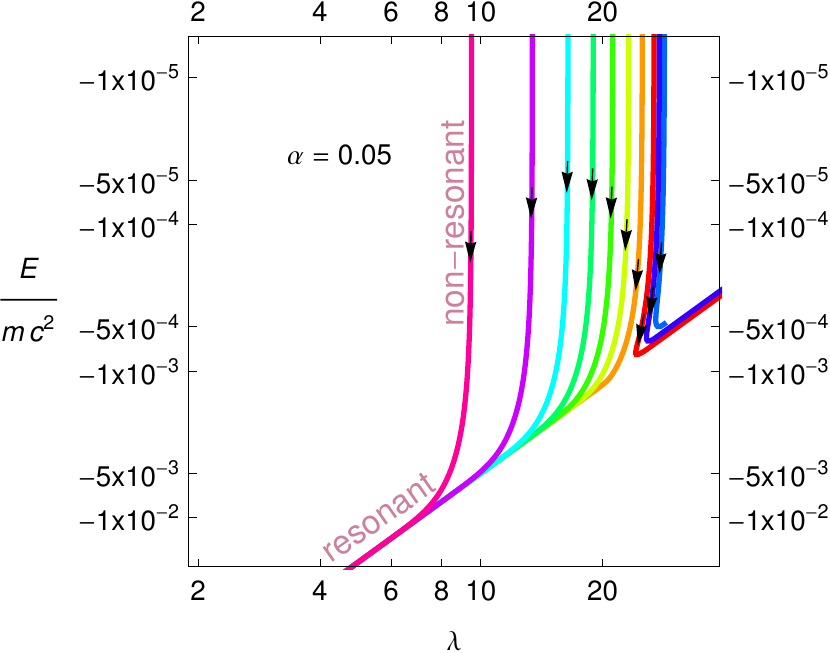}
  \caption{As tides drive the tidal bulge across the surface of the solid asteroid, work deforming the body is done at the expense of orbital energy; little angular momentum is deposited in spinning the body because of its low moment of inertia (non-resonant phase). When the body melts and spends more and more time beyond the Roche radius, it elongates, increases the moment of inertia which allows both orbital angular momentum as well as orbital energy to be transferred to spin (resonant phase).}
\label{Fig:FigEvolution}
\end{figure}

\begin{figure}
\includegraphics[width=\columnwidth]{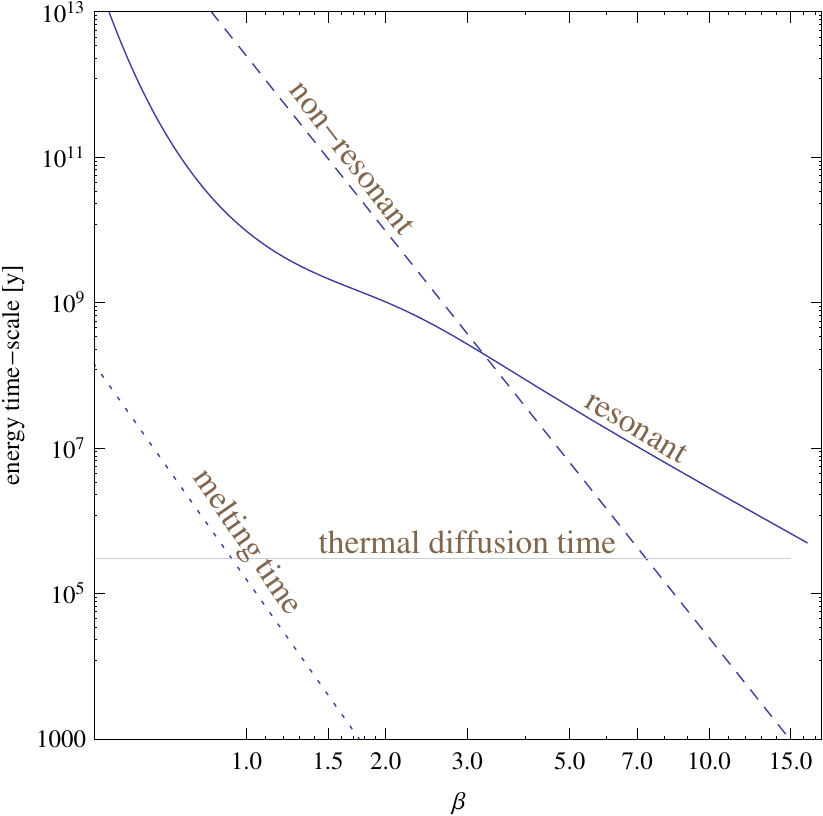}
  \caption{Resonant and non-resonant time-scales as a function of $\beta$ for $\rho = \rho_0$, R $= 10$Km, Q $=$ 200. $\beta$ is the Roche penetration parameter $\beta = r_R/r_p$ and Q is the resonant damping factor.}
\label{Fig:FigResonant}
\end{figure}

Tides induce forced oscillations on the object, exciting its eigenmodes. They are non-resonant, as long as the frequencies $\omega$ of spectral components of tidal interaction are much lower than the quadrupole eigenfrequencies $\omega_q$ of the object. If, on the other hand, spectral components of tides are strong at frequencies close to eigenfrequencies of the quadrupole oscillations of the object, which happens when the orbit touches the Roche radius, tides become resonant. \reff{Fig:FigEvolution} and \reff{Fig:FigResonant} summarize our results.

As indicated in \reff{Fig:FigEvolution}, tidal evolution of the orbit has two distinct phases: the non-resonant phase, well describable with Hut's slow tides formalism and the resonant phase, which was studied numerically in affine approximation by \cite{1986ApJS...61..219L} and in Newtonian approximation by \cite{Gomboc:2005}. The energy loss per periastron passage can in both cases be described by the expression
\begin{equation}
\Delta E = - \left({{G M_{\rm bh} mR^2}\over{r_{\rm
p}^3}}\right)\varepsilon^2(\beta_{eff}) \ ,
\label{tidalwork}
\end{equation}
where $\varepsilon^2(\beta_{eff})$ is the eccentricity of the ellipsoid to which the originally spherical object is deformed by tides; it depends essentially only on the effective Roche penetration parameter $\beta_{eff}=r_{R_{eff}}/r_p$\footnote{This approximation is valid for small  amplitude resonant tides and is valid as long as most part of the orbit is outside the Roche radius, i.e. on highly eccentric orbits whose periastron reaches below Roche radius.}. In the case of gravity dominated objects $\beta_{eff} = \beta = r_R/r_p$ and in the solid case $\beta_{eff}$ is smaller; it can be expressed as $\beta_{eff}=\frac{c_s}{8R}\sqrt{\frac{3\pi}{G\rho} }\frac{r_R}{r_p}\approx 3300\ \mathrm{km}/R\times \beta$, where the factor 3300 comes from inserting the approximate value for the velocity of sound and the density of solid bodies. The time scales for non-resonant and resonant tidal evolution phase can then be calculated (cf. \cite{Cadez:2008}). They can be expressed in the form
\begin{align}
t_{\rm EN} &= 1.4\times 10^{11}yr\times \frac{Q f_{EN}(e)}{\beta^8}\left( \frac{10{\rm km}}{R}\right )^2\left(\frac{\rho_0}{\rho}\right)^{\frac{7}{6}}\\
t_{\rm ER} &= 2.9\times 10^{10}yr\times \frac{ f_{ER}(e)}{\varepsilon^2(\beta)\beta^{\frac{7}{2}}}\left( \frac{10{\rm km}}{R}\right )^2\left(\frac{\rho_0}{\rho}\right)^{\frac{7}{6}}\ ,
\label{ResCas}
\end{align}
where $Q$ is the resonant damping factor, $f_{EN}(e)$ and $f_{ER}(e)$ are factors of order unity depending on the eccentricity of the orbit ($e$), and $\varepsilon^2(\beta)$ is a function that peaks at resonance i.e. at $\beta = 1$.

These time-scales are plotted in \reff{Fig:FigResonant} for 10 km objects with $\rho=5g\ cm^{-3}$ and $Q$=200. Also plotted in this figure is the approximate melting time, i.e. the time it takes to accumulate enough tidal work inside the object to melt it ($t_{melt}=E_{melt}t_{orb}/\Delta E$; $E_{melt}\sim 10^{-10}m\ c^2$), and the thermal diffusion time for such an object (assuming the thermal diffusion coefficient $D\sim 10^{-6}m^2s^{-1}$, characteristic of most rock). It is clear from this figure that 10 km objects melt for $\beta>1$, i.e. if their periastra reach below 33 $r_{\mathrm{g}}$ in less then $3\times 10^5$ years. Asteroids that do not penetrate as deep or are smaller would not melt, because their thermal diffusion time scale is too short\footnote{Note that tidal heating is rather slow, which indicates low tidal stress, so that the break-up of such asteroids is not expected.}. On the other hand asteroids that are scattered or released to highly eccentric orbits with pariastra at 10 $r_{\mathrm{g}}$ melt in only a few years. When melted, asteroids become resonant and their tidal evolution time-scale shortens considerably, certainly to below a few billion years but possibly even to a much shorter time scales, depending on their size and most of all on the periastron of the orbit that they were ejected or scattered to. Increased tidal heating does not essentially raise their temperature, because convection sets in so that low temperature ($\sim$1000 K) black body radiation can dissipate tidal heat. One can conclude that all asteroids found within $\sim 33\ r_{\mathrm{g}}$ are melted, non-evaporated and are bound to end up in the black hole.
\section{The free fall phase}
At present we do not have a complete theory describing all the steps of evolution toward plunging into the black hole. Clearly, at $33\ r_{\mathrm{g}}$ asteroids have entered the resonant part of their tidal evolution in relativistic regime. \reff{Fig:FigEvolution}, based on Hut's slow tide approximation, indicates that the characteristic of this part of evolution is the transfer of orbital angular momentum to spin. In the above resonance regime this transfer can not be effective by spinning the object faster, but by deforming it into a longer and longer thread, thus increasing its moment of inertia. Some understanding of this mechanism can be obtained considering the virial theorem applied on the self-gravitating mass of an orbiting asteroid. In \cite{Gomboc:2005}, the authors consider a self gravitating object moving in the gravitational field of a point mass (Newtonian approximation). Following the steps similar to those in deriving the virial theorem, they obtain the equation
\begin{equation}
W_{int}+\frac{1}{2}W_G=-G\frac{m_{\rm bh}{\bf R\cdot Q\cdot R}}{R^5}+\frac{1}{4}\ddot J\ ,
\label{virial}
\end{equation}
where $W_{int}$ is the internal energy of the asteroid, $W_G$ is the self-gravitational potential energy, ${\bf R}$ is the orbital vector from the black hole to the asteroid, ${\bf Q}$ is the quadrupole tensor of the deformed asteroid, and $J$ is the momentum scalar ($J=\int \rho r^2 dV$). If one approximates the deformed asteroid by an axis-symmetric ellipsoid with the symmetry axis in the direction of $\hat n$, then ${\bf Q}$ can be written in the form ${\bf Q}=q(3\hat n\hat n-{\bf I})$, where ${\bf I}$ is the identity matrix and $q=-\frac{2}{5}M^*(r_{pole}^2-r_{eq}^2)$. Furthermore, $J=\frac{1}{5}(r_{pole}^2+2r_{eq}^2)$, and equation \ref{virial} becomes
\begin{align}
&W_{int}+\frac{1}{2}W_G=\nonumber \\
&\frac{M^*}{5}\left(\omega_K^2(r_{pole}^2-r_{eq}^2)(6\cos^2\alpha-2)+\frac{1}{4}({\ddot{r}^2}_{pole}+2\ddot{r}^2_{eq})\right)\ .
\end{align}
Neglecting the equatorial radius with respect to the polar one ($r_{pole}\gg r_{eq}$), this becomes
\begin{equation}
W_{int}+\frac{1}{2}W_G= \frac{M^*}{5}\left(\omega_K^2 r_{pole}^2(6\cos^2\alpha-2)+\frac{1}{4}{\ddot{r}^2}_{pole}\right)\ ,
\label{virial1}
\end{equation}
where $\alpha $ is the angle between ${\bf R}$ and $\hat{n}$ and $\omega_K=G m_{\rm bh}/R^3$ is the Keplerian angular velocity corresponding to a circular orbit at radius $R$. In stationary equilibrium the right hand side of this equation vanishes and the expression reduces to the well known virial theorem. In the other extreme, if tides had elongated the object a few times its original radius, the left hand side of the above equation becomes negligible with respect to the first term on the right, so that the equation has become a second order linear differential equation for $r^2_{pole}$. Its solutions are oscillatory if the average value of $(6\cos^2\alpha-2)$ is positive, and exponentially expanding if it is negative. For slow tides, the bulge points almost in the direction of ${\bf R}$ ($\alpha \rightarrow 0$) and accordingly $r^2_{pole}$ is an oscillatory function. However, above resonance, the tidal bulge may point more and more in the direction of motion, which changes the character of the above equation making the tidal bulge grow exponentially.

\begin{figure}
  \includegraphics[width=\textwidth]{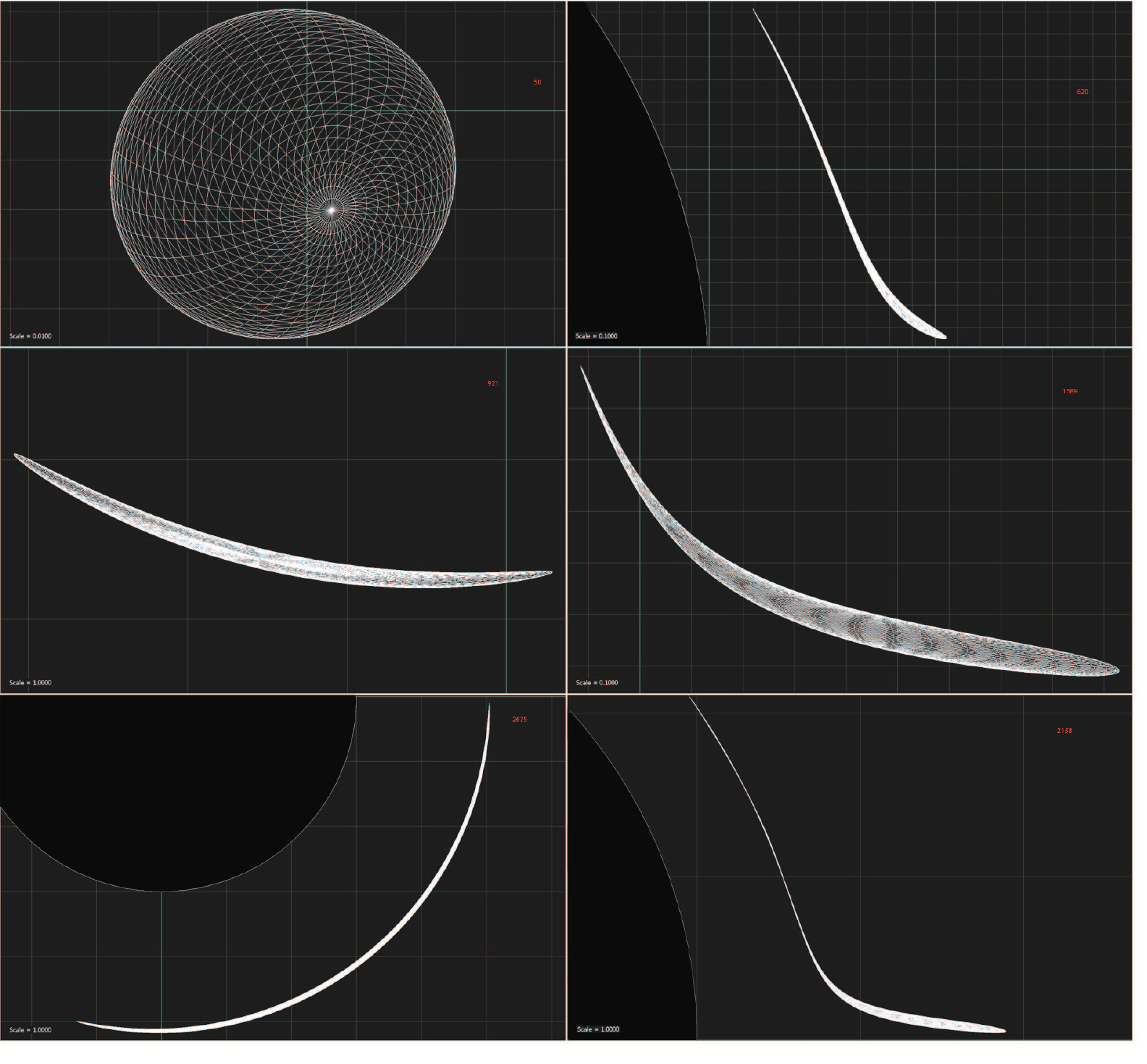}
  \caption{A selection of images from simulations of tidal development of an $0.03\ r_g$ object following an elliptic orbit around a black hole observed from above the orbital plane (frames on the left) and as seen from $1^\circ$ above the orbital plane. Captions to left of frames indicate the size of grid spacings; green grid lines mark coordinates which are integers of $r_{\mathrm{g}}$. Effects of gravitational lensing are clearly seen in frames on the right.}
\label{Fig:Slides}
\end{figure}

\begin{figure}
\includegraphics[width=\columnwidth]{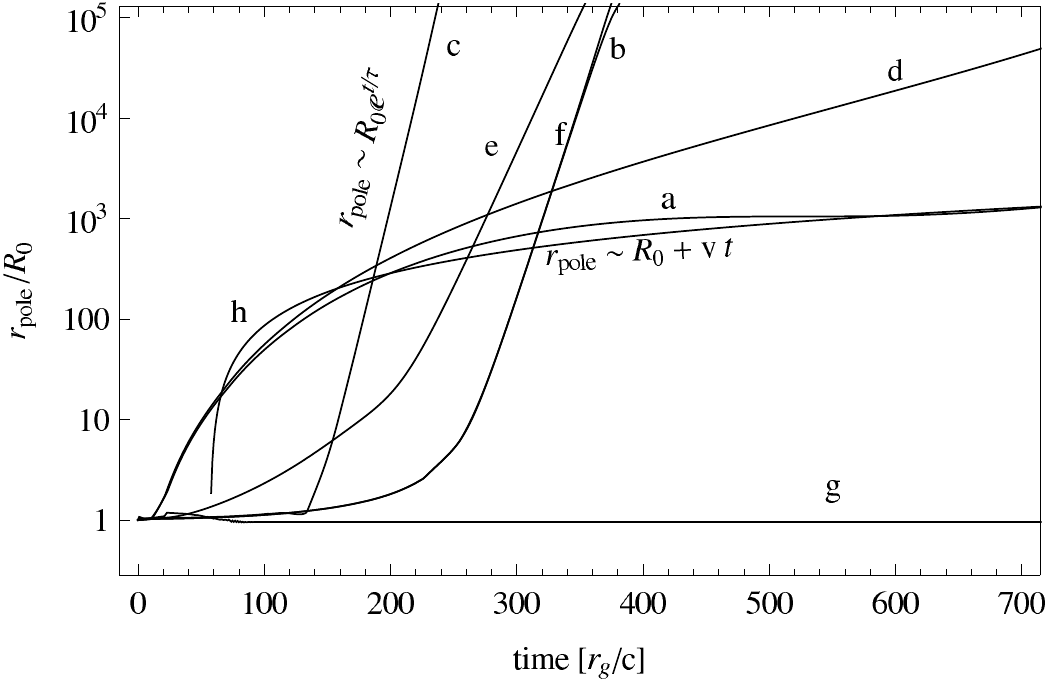}
  \caption{The length $2r_{pole}$ of the image of a tidally-evolving object seen above the orbital plane as a function of time for different orbital parameters: a) elliptic orbit close to innermost stable circular orbit (ISCO) with $r_{\rm A} = 6.2122\ r_{\rm g}$, $r_{\rm P}=6.2097\ r_{\rm g}$ and $\zeta=2.44\times 10^{-5}$; b) plunging parabolic orbit with $E/mc^2 = 1$, $l/mr_{\rm g}c= 3.999998$ and $\zeta=1+10^{-6}$; c) plunging hyperbolic orbit with $r(V_{\mathrm{max}}) = 3.6\ r_{\rm g}$ and $\zeta=1+10^{-6}$; d) elliptic orbit close to ISCO with $r_{\rm A} = 6.1\ r_{\rm g}$, $r_{\rm P} = 5.95\ r_{\rm g}$ and $\zeta=1-3\times 10^{-6}$; e) elliptic orbit with $r_{\rm A} = 20\ r_{\rm g}$ , $r_{\rm P} = 4.45\ r_{\rm g}$ and $\zeta=1-3\times 10^{-6}$; f) scattering parabolic orbit with $E/mc^2 = 1$,  $l/mr_{\rm g}c= 4.000002$ and $\zeta=1-3\times 10^{-6}$; g) plunging hyperbolic orbit with $E/mc^2=3.0$ and $l/mr_{\rm g}c=6$ and $\zeta=6.35$; h) linearly growing deformation. In cases b,c,e and f, the simulation was halted before the object evolved beyond a limit of numerical accuracy.}
\label{Fig:ExponentialGrowth}
\end{figure}

The virial theorem is not sufficient to describe this phenomenon in any detail, because it does not predict the value of the angle $\alpha$. Therefore, we study the last stages numerically with a fully relativistic code. This code, described in \cite{Kostic:2009} builds an image of the tidally evolving object assuming that it starts as a molten sphere. In calculating the developing shape of the body fully relativistic dynamics is taken into account, with only one simplification that the internal pressure can be neglected with respect to the high kinetic energy density of flow in the tidal bulge. Evolving images of the object are constructed as they would be seen momentarily in the frame of a distant observer. A selection of a few slides in \reff{Fig:Slides} shows such a development and confirms predictions of exponential growth in the above simple argument. In \reff{Fig:ExponentialGrowth} we also show the development of the visible length of an originally spherical object as it is stretched by tides, obtained in a number of numerical experiments. We find that resonant tides develop exponentially growing tidal tails which are fastest and most pronounced in the vicinity of the black hole.

\begin{figure}
\includegraphics[width=\columnwidth]{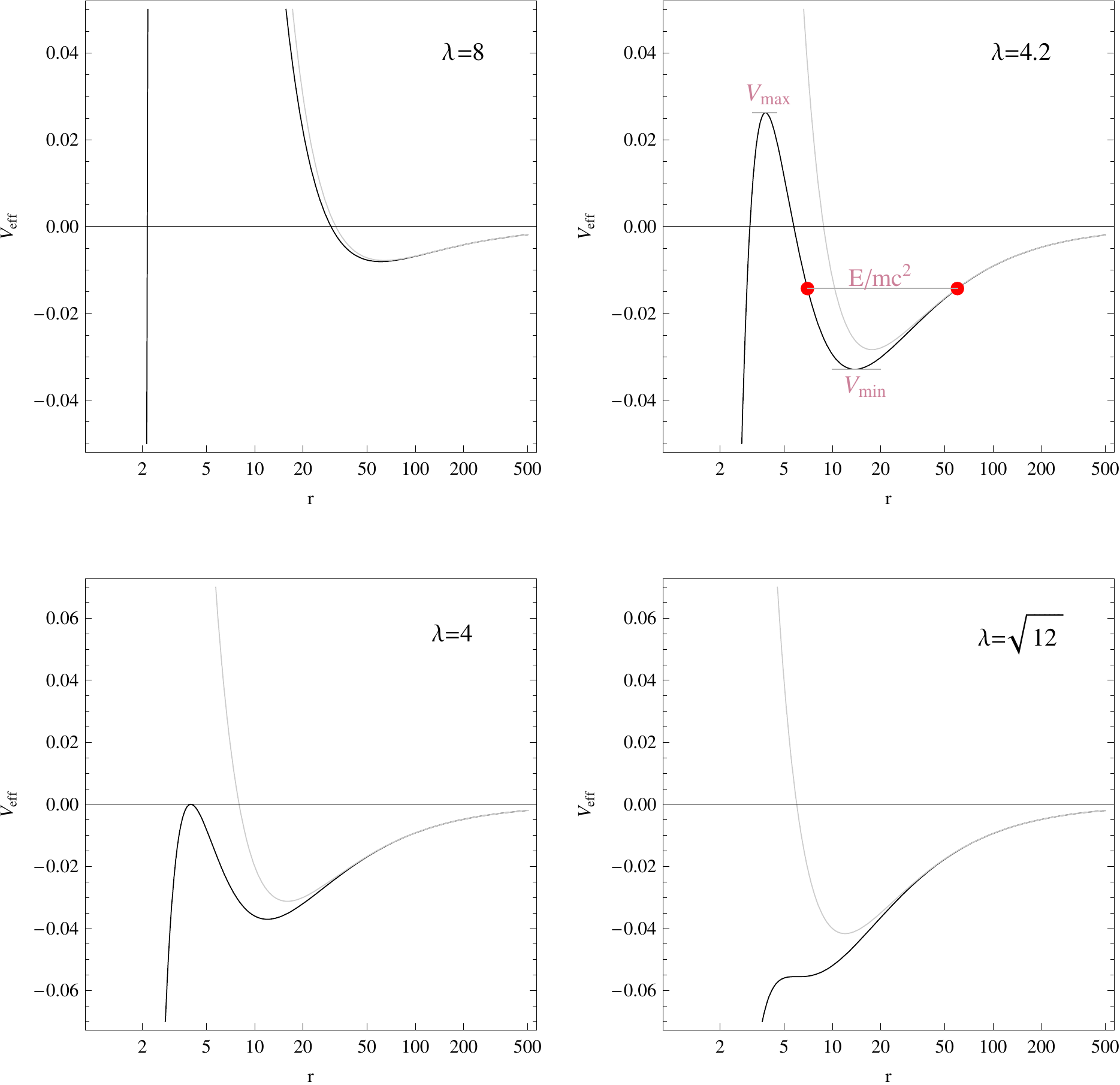}
  \caption{The Schwarzschild (black) and Kepler (light gray) effective potentials for different values of dimensionless angular momentum ($\lambda=\frac{l}{m\ r_g c}$)}
\label{Fig:pot}
\end{figure}

At this point it is important to appreciate the difference between different tidal evolution problems. In the Keplerian case, described so well by Hut's mechanism, the tidal dissipation of orbital energy and exchange of angular momentum between orbits and spins brings orbital parameters closer and closer to equilibrium at the minimum of the effective potential. This equilibrium may be reached (depending on initial conditions) either at a distance where the pair is well separated so that no partner is within each others Roche radius, or on an orbit where one partner finds itself beyond the Roche radius of the mutual potential. In the first case, tidal evolution has stopped since tides can no more dissipate energy (the only dissipation mechanism left is by gravitational radiation which is much slower in most cases). In the other case, the member that finds itself beyond its Roche radius spills over to the companion by building an accretion disk around it. Unless there is a strong interaction with other bodies in the vicinity, the accretion phase develops, according to Hut's mechanism, during a stage when the members of the binary are (almost) co-rotating. Therefore, according to the above discussion, the angle $\alpha$ remains close to zero, keeping equation \ref{virial1} in oscillatory regime. This ensures that the shape of the mass-losing star remains stable. Mass flow through the inner Lagrangian point is made possible only by the (slow) transfer of angular momentum enabled by energy dissipation in the accretion disk. If, on the other hand, the accreting mass is a black hole and the inner Lagrangian point is close to the maximum of the black hole effective potential, then most geodesic orbits from the mass-losing companion lead directly behind the horizon of the black hole. No stationary accretion disk can form in this case and the mass-losing companion is sucked to the black hole in one gulp as a characteristic exponentially elongating string. In our numerical experiments, we found that the dividing line between non-relativistic and relativistic accretion is well represented by a relativistic parameter $\zeta$
\beq
\zeta=\frac{E/mc^2-V_{min}}{V_{max}-V_{min}}\ , \label{eq:zeta}
\eq
where $E$ is the orbital energy of the asteroid and $V_{max}$ and $V_{min}$ are the maximum and minimum of the relativistic effective potential. Relativistic accretion occurs only if $\zeta \sim 1$. Even orbits winding beyond the Roche radius and close to the black hole, but with orbital energy near to $V_{min}$, do not accrete relativistically. The elongation of the object does not develop exponentially but linearly, as shown by the curve 'a' in \reff{Fig:ExponentialGrowth}. The meaning of $\zeta$ can be visualized in \reff{Fig:pot}. It is clear from this figure that objects on bound orbits accrete relativistically only if their orbital angular momentum decreases below $4 m r_g c$. Thus, relativistic accretion can occur only from orbits with periastra between 4 and 6 $r_g$. Such unstable relativistic orbits wind about the black hole few times and unwind from there into the black hole. The winding periods ($T=\frac{1}{c}\sqrt{4\pi^2 r_p^3/r_g}$) on such orbits are between 15 and 28 minutes (for a $3.6\times 10^6M_\odot$ GC black hole).

The main conclusion of this section is the realization that tidal evolution of binary systems with a black hole has three, and not only two possible outcomes as in Newtonian theory. Namely, accretion of a smaller object onto a black hole would not occur by gradual Roche lobe overflow as in a contact binary, but in a single relativistic gulp, if the Roche lobe overflow has to occur at the inner Lagrangian point between 4 and 6$r_g$ from the black hole. The energy released during such an event is probably comparable or more then the energy released during the whole disk accretion phase and the light-curve is modulated with the last period of revolution about the black hole. The new relativistic outcome occurs, if the binary started as detached and the Roche lobe overflow occurs suddenly (due to melting) near the periastron which is below $6r_g$.
\section{The role of magnetic fields}
NIR flaring emission from $\sgra$ is believed to be produced by synchrotron radiation of electrons with Lorentz factors $\gamma_e \sim 10^2 - 10^3$ moving in a magnetic field of $10-100\ \mathrm{G}$ \cite{Markoff:2001, Baganoff:2001, Yuan:2004}. In this section we would like to show that the development of such strong magnetic fields may be a natural development during the short event of relativistic accretion. We would also like to show, in a qualitative manner, that if certain physical conditions are met, then the global magnetic (and electric) field may play the decisive role in changing the large kinetic energy of the exponentially extending object into kinetic energy of electrons and finally into radiation, released just before the object forever disappears behind the horizon of the black hole.

There is still no general agreement on the strength of the magnetic field at the Galactic centre. In \cite{LaRosa:2006} it is  pointed out that the field may be strong (of order $1\ \mathrm{mG}$) and globally organized (see e.g. \cite{Morris:2006}) or globally weak (of order $10\ \mathrm{\mu G}$) with regions where it is stronger (see e.g. \cite{LaRosa:2005, Boldyrev:2006}). In any case, the observed strength is orders of magnitude lower than requested by synchrotron models. Assuming that the object is permeated with such low magnetic field, our numerical simulations suggest a simple mechanism to increase it, as well as to provide electrons with high $\gamma_e$.

We have found that objects moving along $\zeta\sim 1$ orbits end up being exponentially squeezed and elongated. Since the inward squeezing velocity is low ($v/c\sim 10^{-7}$), it is possible to use the magneto-hydrodynamic (MHD) approximation to describe the evolution of the magnetic field
\beq \frac{1}{\sigma \mu_0}\nabla\times(\nabla\times\vec{B}) =
-\parc{\vec{B}}{t} + \nabla\times(\vec{v}\times\vec{B})\ ,
\label{eq:induction}
\eq
where $\sigma$ is the electrical conductivity and $\vec v$ is the plasma velocity. Assuming typical values $\sigma=4.14 \times 10^6\ \Omega^{-1}\mathrm{m}^{-1}$ for molten aluminium and $L=100\ \mathrm{km}$ for the size of the object, the magnetic Reynold's number $R_m$ becomes very large
\begin{displaymath}
R_m = \mu_0 \sigma v L \approx 10^7\ .
\end{displaymath}
Therefore we may neglect the diffusion term in the above equation
\beq
\parc{\vec{B}}{t} = \nabla\times(\vec{v}\times\vec{B})\ ,
\label{eq:induction_frozen}
\eq
which, according to Alfv\' en's frozen-flux theorem, leads to magnetic flux conservation. Consider an object with initial cross section $S_0$ which is permeated with the initial magnetic field $B_0$. If such an object is exponentially squeezed, magnetic flux conservation increases the magnetic field density as $B=B_0 S_0/S$. Since the tidal deformation tensor is traceless, the average cross section area of a squeezed and elongated object is inversely proportional to its length (in the GR case this result is only approximate). Therefore, the magnetic field at a later times can be estimated as
\beq
B(t) \sim B_0 \frac{r_{pole}(t)}{R_0} \ ,
\label{eq:B}
\eq
where $R_0$ is the initial radius of the object and $r_{pole}(t)$ the major axis of the ellipsoid, to which the object has been deformed at time $t$. In \reff{Fig:ExponentialGrowth}, results of dynamical simulations show length increase to follow the exponential law up to five orders of magnitude. The limit comes from numerical limitations of our simulation only - at this factor of expansion the numerical grid became too coarse to trust further calculations. In fact, in the approximation taken, where internal pressure was neglected, (almost) no limit to the expansion factor is expected. Equation \ref{eq:B} obviously indicates that neglecting he internal pressure has its limits. When the magnetic pressure becomes comparable to dynamical pressure involved in driving the expanding flow, the increase of magnetic pressure must be stopped, impeding further contraction.

In order to estimate dynamical pressure inside the exponentially expanding object, we consider a simplified problem with a solution that is similar to the exponentially expanding asteroid solution from previous section. Consider a spherical self-gravitating incompressible fluid that suddenly finds itself in a strong field of a cilindrically symmetric gravitational quadrupole ($\phi_g=\frac{r_gc^2}{r^3}(x^2+y^2-2z^2)$). An exact solution of this hydrodynamic problem can be found such that the sphere is continually deformed into and exponentially extending ellipsoid with axes $r_{pole}(t)=R_0 e^{t/\tau}$ and $r_{eq}(t)=R_0 e^{-t/2\tau}$, with $\tau=(2 r_g/r^3)^{-1/2}/c $, and the central pressure decreases (for $t\ \ni$ $r_{pole}\gg r_{eq}$) as
\begin{align}
p_c&=\frac{3}{4}\rho\ c^2 \frac{r_g}{r}\left (\frac{r_{eq}(t)}{r}\right )^2\nonumber \\ 
&=\ \frac{3}{4}\frac{\rho r^2_{eq}(t)}{\tau^2}
\label{cntralP}
\end{align}
Note that the central pressure is 3/4 of what it would be in hydrostatic equilibrium. If magnetic pressure pushing outward at equator is added and balances the central pressure, there is no more pressure gradient toward equator, so radial compression must stop. The negative magnetic pressure in the axial direction is also working against expansion, however the expansion velocity along this axis is much larger than the compression velocity, so expansion can only be halted if the kinetic energy of expansion is transformed into some other form, as for example magnetic energy or kinetic energy of electrons. Equating the central hydrodynamic pressure and the magnetic pressure $p_m=\frac{1}{2\mu_0}B^2$, one obtains the equation for the equatorial radius when compression stops ($R_{eq}$)
\begin{equation}
\frac{R_{eq}}{R_0}=\left(\frac{2}{3\mu_0 \rho}\right)^{1/6}\left(\frac{\tau}{ R_0}\right )^{1/3}B_0^{1/3}\ .
\label{pressBal}
\end{equation}
Inserting the numbers used in previous arguments ($\tau\sim 11r_g\sim 200 {\rm sec}$, $\rho=5 g\ cm^{-3}$) and measure $R_0$ in units of 10 km and $B_0$ in units of {\rm mG}, then the above becomes
\begin{equation}
\frac{R_{eq}}{R_0}=0.0027\times \left(\frac{B_0[{\rm mG}]}{ R_0[10{\rm km}]}\right )^{1/3}\ ,
\label{presE}
\end{equation}
and the ratio of polar ($R_{pole}$) to initial radius follows as
\begin{equation}
\frac{R_{pole}}{R_0}=133000\times \left(\frac{B_0[{\rm mG}]}{ R_0[10{\rm km}]}\right )^{1/3}\ .
\label{presP}
\end{equation}
Equations \ref{presE} and \ref{presP} tell that a 10 km asteroid permeated with an initial magnetic field of 10 {\rm mG} stops compressing when its equatorial radius is 27 meters, the magnetic field has increased to 133{\rm G}, the polar radius extended to $1.33\times 10^6{\rm km}$, and the whole event lasted $\sim 40 min$ - less than 3 turns around the black hole. The final velocity of the pole with respect to the centre of mass is about $6.6\times 10^3{\rm km/sec}$ and the kinetic energy of expansion of the order $2\times 10^{-4}m\ c^2$.

These 40 minutes are rather eventful. Very few processes in the universe release so much energy so fast. The work of gravity on the still extending object can go on to extract in a few more turns up to $ \sim 0.1 mc^2$ in tidal energy (\cite{Gomboc:2005}). 

This mechanism works as long as the magnetic dissipation time-scale $\tau_{\mathrm{D}}=\sigma\mu_0 r^2_{eq}$ is longer than the dynamical time-scale. Initially $\tau_{\mathrm{D}}/sim 1.2\times 10^8 {\rm sec}$ and this condition is certainly well satisfied. Finally, when the object is only 27 meters in diameter, the value of $\tau_{\mathrm{D}}$ drops to $\sim 15{\rm min}$ if the same molten metal conductivity is assumed. It is quite clear that this assumption can no longer be respected, since after the release of so much energy, the object can no longer be considered a molten metal but has obviously turned into hot plasma. The conductivity probably increases and keeps magnetic flux constant.

As long as the magnetic field is low, it follows the motion of the plasma. However, when it reaches high enough values, matter starts to move according to this magnetic field, thus providing a mechanism for accelerating electrons to relativistic velocities. The gravitational force which guides the motion of positive charges is negligible for electrons. They are pushed out by magnetic pressure until charge separation induces enough electric field to keep them bound to positive charges beneath. The electrons that are pushed out spiral along magnetic field lines and experience less scattering by heavy ions. Eventually, they lose energy mostly by synchrotron radiation. As the electrons gradually gain speed also in the direction of the magnetic field lines, the polarization of the emitted radiation also changes.

Of course, only a complete MHD simulation can give precise results, showing the geometry of currents responsible for such a magnetic field, and could predict precisely how much energy could be transformed into radiation, what would be the spectrum and how the polarization would change with time. The above arguments may only give us some confidence that such processes actually can occur at the Galactic centre.
\section{A simple scenario}
Let us propose the following scenario to explain flares from the Galactic centre. Let us consider an object (LMS) on a highly eccentric orbit with periastron below $\sim 10\ \rg$, that got there by non-resonant relaxation. The tidal force melts the LMS and eventually brings it on a $\zeta\sim 1$ critical orbit. Just for the sake of simplicity, let us assume that the critical orbit is a parabolic one, with the critical radius of $4\ \rg$. Let us assume moreover that the energy released during this process comes from the gravitational potential energy of the object and is thus proportional to its mass. Since potential energy differences on an orbit at $r=4\ r_{\rm g}$ are of the order of a few percent of $mc^2$, it follows that objects producing the flares probably have masses of the order of $10^{20}\ \mathrm{g}$. If the sources of flares were gaseous blobs of such a large mass, they would find themselves below the Roche radius far away from the black hole, and would therefore be completely disrupted before producing any modulation of the light curve. We conclude that the source of the flares is a small and initially solid object that orbits the black hole above the effective Roche radius, and is then brought close to the black hole, where it melts.

As the tidal deformations on this orbit grow exponentially with time, the same happens with the magnetic field and the magnetic pressure, which resist the tidal squeezing and stretching -- the magnetic field helps preventing the object from falling apart. The accelerations of the electrons are the highest at the endpoints of such a thread, because the magnetic field is highly inhomogeneous there. Consequently, most of the synchrotron radiation is emitted from that region.

Since the tidal deformation tensor is traceless and the magnetic pressure tensor is not, the tidal squeezing of the object is stopped by magnetic pressure, while the tidal stretching along the orbit continues. Eventually, due to this imbalance, the object breaks into a large number of smaller pieces, just as in case of Plateau-Rayleigh instability. Every new piece emits most of the radiation at its endpoints. These smaller pieces move on the same orbit, because the magnetic field prevents them from completely spreading around the black hole.

The light-curve of such an event can be constructed in the following way: the small, radiation emitting pieces are represented by $N$ small rigid objects, orbiting on a critical orbit with the critical radius of $4\ \rg$. The orbit is chosen in such a way, to match the time-scales with the observations, i.e.\ $50\ \rg/c$. Each of the $N$ objects has the same light-curve with a different phase shift. The objects which are farther away from the black hole are less deformed, have lower magnetic field and thus lower luminosity with respect to the ones closer to the black hole. Due to exponentially increasing tidal deformation, the light-curves of all the objects are modified by an exponential function, to reflect this. Finally, all the modified light-curves are integrated into one.

The appearance and the luminosity of this event is calculated with respect to the observers at different inclinations and position angles with respect to the orbit of the in-falling object using ray tracing techniques. Our simple scenario has five parameters: the mass of the black hole $\mbh$, the tidal ``squeezing'' time scale $\tau_{\rm h}$, the length of the tidal tail $l_{\rm tail}$, the inclination of the orbit $i$ and the longitude of the line of nodes $\Omega$. 
\begin{figure}
  \includegraphics[width=\columnwidth]{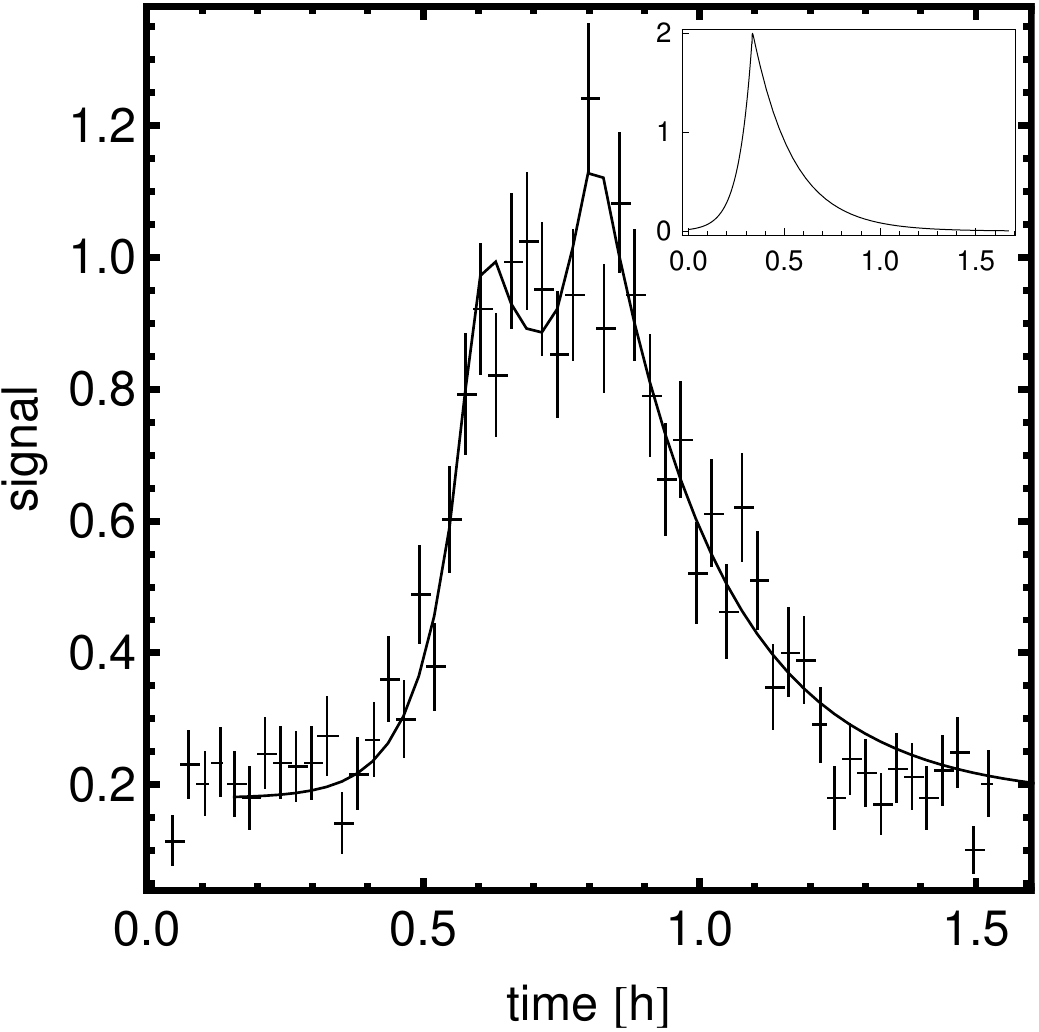}
  \caption{XMM-Newton/EPIC flare of April 4, 2007, fitted with the above simple model. The inset shows (roughly) the intensity at a fixed position of the image as a function of time; the rising beginning corresponds to the ``squeezing'' time-scale and the duration of the falling edge corresponds to the length of the tidal tail.}
\label{Fig:flare}
\end{figure}
An example of such a simple model is a fit to an observed Galactic flare light-curve (flare of April 4,2007, \cite{2008A&A...488..549P}), shown in \reff{Fig:flare}. This light-curve is consistent with the following parameters (assuming a Schwarzschild black hole at GC): $m_{bh}=2.9\times 10^6M_\odot$, $\tau_{\rm h}=4.8{\rm min}$, $l_{\rm tail}=6r_g$, inclination near $90^{\circ}$ and $\Omega$ is undetermined, since there are only two peaks in the light curve. The mass following from the fit does not quite agree with the measured one. The discrepancy can easily be explained by introducing a Kerr black hole with angular momentum parameter 0.4. However we prefer not to dwell on such details before a more detailed understanding is reached.
\section{Conclusions}
Given for granted that our Galactic centre harbours a super-massive black hole, one of its puzzling characteristics are the flares detected at various wavelengths. Their main characteristics are: a) short duration; b) quasi-periodic oscillations with the period of 17-22 minutes; c) total energy release of the order $10^{39.5}\ {\rm erg}$; d) strong indications of the presence of magnetic fields up to 100 gauss.

The exact cause of such flares is still unclear, and several models have been suggested to explain their origin. We have proposed and explored the idea that the flares are produced by the final accretion of dense objects (LMS) with mass of the order of $10^{20}\ {\rm g}$ onto the massive black hole in SgrA$^*$. To this aim, we have investigated the effects of black hole tides on small solid objects in the vicinity of a massive black hole. Such objects are expected to populate the Galactic centre as a result of being stripped off their parent stars. In \cite{Cadez:2008}, we have shown that solid objects, in the mass range $\sim 10^{19}-10^{21}\ {\rm g}$, melt in the vicinity of the black hole. Their subsequent orbital evolution naturally leads to capture orbits, most likely via unstable circular orbits. In an exhaustive numerical simulation we studied the evolution of shapes and light-curves as the objects progress toward the horizon of the black hole. We find that tides generally elongate objects into long thin threads, which extend exponentially on a time scale $\tau_c\approx 11.3\ r_{\rm g}/c$, if the orbital criticality parameter $\zeta$ is close to 1. Moreover,if the object is bigger than $\gtrsim 0.3\ r_{\rm g}$, it is completely disrupted before producing any significant modulation of the light curve. Therefore the object should be relatively small to produce the observed modulation.

We studied numerically the tidal evolution of the shapes and the images of such objects in the gravitational field  of a Schwarzschild black hole. Because of the complexity of the problem, we did not consider the more general Kerr space-time and limited the analysis to pressure-less gas. We find, quite generally, that such objects are eventually elongated into thin threads, which on $\zeta\sim 1$ orbits extend exponentially.

We point out that during such exponential elongation, conditions for magnetic flux conservation are met, so that relatively large magnetic fields can be generated. The exponential growth of magnetic field density also allows the betatron mechanism which can generate highly relativistic electrons. We propose this process as a possible source of galactic flares. We do not have the exact mechanism to extract radiation from gravitational potential energy, yet we are able to model light curves of galactic flares with simple assumptions.

Finally, we would like to point out that our scenario differs from other models in the following respects: 1) Flares originate from marginally bound and not from marginally stable orbits. This brings them closer to the black hole and provides a mechanism for fast extraction of energy. It also agrees with recent measurements of the size of flaring region in the Galactic centre, showing that it is not larger than $6\ r_{\rm g}$, see e.g. \cite{Reid:2008}, and relaxes constraints on the angular momentum of the black hole. 2) Our model does not require a fairly rapidly rotating black hole as needed e.g. by \cite{Genzel:2003, Aschenbach:2004, Trippe:2007}. 3) An accretion disk at the Galactic centre is not needed. In \cite{Yusef-Zadeh:2008} it is highlighted that ``there is no definitive evidence that \mbox{Sgr~$\mathrm{A}^*$}\ has a disk''. 4) Elongated structures invoked in other models to fit flare data, e.g. \cite{Falanga:2007, Zamaninasab:2008, Hamaus:2008}, develop naturally in the proposed model as a result of the tidal evolution of a melted body. 5) The evolving elongated structures also provide a natural mechanism for fast changes in magnetic field density inferred from observations, increasing first as the body is squeezed by tides and finally decreasing as the body is crossing the horizon of the black hole.
%
%
%
\begin{theacknowledgments}
  A.\v C. would like to thank the organizers and in particular to prof. Ruffini for the invitation to take part in the lively meeting honouring the late prof. Zeldovich.   
\end{theacknowledgments}
%
%

%
\end{document}